\documentclass[12pt]{iopart}
\usepackage{graphicx}
\usepackage{subfigure}
\usepackage{epsfig}
\usepackage{epstopdf}

\usepackage{iopams}

\begin{document}

\title{Maximizing optomechanical entanglement with optimal control}

\author{Dionisis Stefanatos}

\address{Hellenic Army Academy, Vari, Athens 16673, Greece}
\ead{dionisis@post.harvard.edu}
\vspace{10pt}
\begin{indented}
\item[]October 2016
\end{indented}

\begin{abstract}

In this article, we formulate the generation of optomechanical entanglement between the linearly coupled cavity field and the mechanical resonator as an optimal control problem in hyperbolic space $H^3$, with control input the coupling rate of the two oscillators. Next, we use optimal control theory to find the allowed optimal values of the coupling which maximize the amount of generated entanglement for a fixed duration of the interaction. Finally, we employ a numerical optimization method to obtain the exact optimal pulse sequences for several illustrative examples. In the strong coupling regime, where the coupling rate is comparable or larger than the frequency of the mechanical resonator, a substantial amount of entanglement can be generated within a fraction of a single oscillator period.

\end{abstract}

%
\vspace{2pc}

\section{Introduction}

Cavity optomechanics, the field of physics studying the interactions of light confined in an optical cavity with the mechanical resonator forming the movable end of the cavity, is a very active area of research with many important applications \cite{Aspelmeyer14}. These include quantum information processing, for the interconversion between storage and communication qubits, high-precision measurements of tiny masses, displacements and forces, and fundamental tests of quantum physics at a macroscopic scale. One fascinating application is the creation of entanglement between the electromagnetic field trapped in the cavity and the motion of the macroscopic mechanical resonator. Several theoretical works have suggested the creation of this optomechanical entanglement in the so-called \emph{steady state} regime, see for example \cite{Vitali07,Paternostro07}. In these and all the related studies, a continuous-wave light field is applied to the optomechanical system and drives it to a steady state, where the cavity filed and the mechanical motion of the oscillator are entangled.

The major drawback of the steady state schemes is that the amount of generated entanglement is limited by the requirement of working in a stable stationary state. In order to overcome this limitation, an alternative \emph{pulsed} approach was suggested \cite{Hofer11}. According to this protocol, the entanglement is not created with the application of a continuous-wave field but with a light pulse, avoiding thus the limits imposed by stability requirements. This pulsed scheme was implemented experimentally for generating entanglement between mechanical motion and the microwave radiation field \cite{Palomaki13}.

In the current paper we also consider a pulsed scheme, where the coupling strength between the cavity filed and the mechanical resonator is modulated in order to create the optomechanical entanglement. This is obviously different than the pulsed protocol discussed above. Another important difference is that here we concentrate in the \emph{strong coupling} regime, where the coupling strength is comparable or larger than the frequency of the mechanical oscillator, while the pulsed protocol \cite{Hofer11} focuses in the week coupling regime (coupling much lower than the mechanical frequency), where the rotating wave approximation can be employed. An important consequence of working with large coupling is that a substantial amount of entanglement can be created within a fraction of the period of the mechanical motion. Note that the pulsed modulation of the coupling has been mostly used for the numerical optimization of optomechanical cooling \cite{Jacobs11,Machnes12,Stefanatos16,Jacobs16}. Recently, it was also employed to numerically investigate the fast entanglement creation between two oscillators (for example two mechanical modes) coupled via their interactions with a third oscillator (optical mode) \cite{Jacobs16}. Here we also formulate entanglement generation as an optimal control problem, but before recoursing to numerical optimization we use optimal control theory to draw as much information as possible for the optimal solution that maximizes entanglement.

The present article is formulated as follows. In the next section we present some basic facts about optomechanical entanglement and identify the target state, while in section \ref{sec:problem} we formulate entanglement generation as an optimal control problem in hyperbolic space $H^3$. In section \ref{sec:solution} we use optimal control theory to study the resultant problem and obtain the allowed optimal values of the coupling strength. In section \ref{sec:examples} we use a numerical optimization method to find the exact optimal pulse sequences for some illustrative examples, while section \ref{sec:conclusion} concludes this work.

\section{Optomechanical entanglement}

\label{sec:entanglement}

In a laser driven optical cavity, the light from the laser entering the cavity is reflected between the mirrors and the enhanced electromagnetic field built inside the cavity exerts a force to the movable mirror due to radiation pressure. The interaction between the cavity (photon) field and the mechanical vibration of the mirror (phonon field) can be described by the following Hamiltonian \cite{Aspelmeyer14}, sufficient for the study of many interesting phenomena in cavity optomechanics,
\begin{equation}\label{hamiltonianfull}
  \mathcal{H}=-\hbar\Delta\hat{a}^+\hat{a}+\hbar\omega_m\hat{b}^+\hat{b}+\hbar \tilde{g}(t)(\hat{a}^++\hat{a})(\hat{b}^++\hat{b}),
\end{equation}
where $\hbar$ is Planck's constant, $\Delta$ is roughly the laser detuning from the cavity frequency and $\omega_m$ is the frequency of the mechanical resonator. The operators $\hat{a},\hat{b}$ are the annihilation operators for the photon and the phonon fields, respectively, while $\hat{a}^+,\hat{b}^+$ are the corresponding creation operators.
The first two terms in (\ref{hamiltonianfull}) express the energy of each individual oscillator (optical and mechanical), while the last term the interaction between them. The coupling rate $\tilde{g}(t)$ has dimensions of frequency and is the available control which can be altered with time, with the aim to entangle the mechanical motion of the mirror with the cavity field (optomechanical entanglement).

The state of the system can be described by the density matrix $\rho(t)$, which satisfies the Liouville-von Neumann equation \cite{Merzbacher97}
\begin{equation}\label{Liouvillefull}
  \dot{\rho}=-i[\mathcal{H}/\hbar,\rho].
\end{equation}
Note that here we consider only the coherent evolution and ignore relaxation, which describes the undesirable interaction of the system with its environment, since we are interested in the fast pulsed regime, where the evolution takes place within a fraction of a single oscillation period, and not the steady state regime. This is a legitimate practice in minimum-time quantum control problems \cite{Stefanatos16,Jacobs16}, the alternative being to include relaxation and maximize the fidelity of the final point \cite{Jacobs11}. We will concentrate on the so-called ``blue-detuned regime" $\Delta=\omega_m$, the case which is more relevant for entanglement generation as we immediately explain. Observe that the interaction term in (\ref{hamiltonianfull}) can be decomposed into two terms as follows
\begin{equation*}
(\hat{a}^++\hat{a})(\hat{b}^++\hat{b})=(\hat{a}^+\hat{b}+\hat{a}\hat{b}^+)+(\hat{a}^+\hat{b}^++\hat{a}\hat{b}).
\end{equation*}
The first term is suitable for optomechanical cooling and state swapping, while the second one for entanglement generation \cite{Hofer11}. For a blue-detuned laser, the second term is resonant with the cavity, and this is why we concentrate on this case for studying the generation of optomechanical entanglement. If we normalize time using the frequency $\omega_m$, i.e. set $t_{\mbox{new}}=\omega_mt_{\mbox{old}}$, then the Liouville-von Neumann equation takes the following form, with $g(t)=\tilde{g}(t)/\omega_m$
\begin{equation}\label{Liouville}
  \dot{\rho}=-i[H,\rho],\;H=-\hat{a}^+\hat{a}+\hat{b}^+\hat{b}+g(t)(\hat{a}^++\hat{a})(\hat{b}^++\hat{b}).
\end{equation}

We consider that at $t=0$ the quantum system starts from the vacuum state
\begin{equation}\label{initial}
|\psi(0)\rangle=|0\rangle_a|0\rangle_b.
\end{equation}
The entangled state which we would like to generate at the final time $t=T$ is closely related to the well known two-mode squeezed vacuum state \cite{Lvovsky15}
\begin{equation}\label{squeezed}
|\phi\rangle=e^{r(\hat{a}^+\hat{b}^+-\hat{a}\hat{b})}|0\rangle_a|0\rangle_b=\sum_{n=0}^{\infty}\frac{1}{\cosh{r}}\tanh^n{r}|n\rangle_a|n\rangle_b,
\end{equation}
where $r\geq 0$ is the squeezing parameter. Note that the above expansion in terms of the number states can be obtained following the procedure described in \cite{Lvovsky15}. Before turning our attention to the target state that we actually  consider in this article, we present some characteristics of state (\ref{squeezed}) which we will need in the analysis that follows.

First of all note that, since (\ref{squeezed}) is a pure state, its entanglement is quantified by the von Neumann entropy of the reduced state \cite{Enk99}
\begin{equation}\label{entanglement}
S=-\sum_{n=0}^\infty p_n\ln{p_n}=\cosh^2{r}\ln(\cosh^2{r})-\sinh^2{r}\ln(\sinh^2{r}),
\end{equation}
where the probabilities
\begin{equation*}
p_n=\left(\frac{\tanh^n{r}}{\cosh{r}}\right)^2
\end{equation*}
are taken from the expansion (\ref{squeezed}) and the final result is obtained by summing the series appropriately. If we set $x=\sinh^2{r}$, then
\begin{equation*}
S(x)=(1+x)\ln{(1+x)}-x\ln{x},
\end{equation*}
which is an increasing function of $x>0$. But $x$ is also an increasing function for $r\geq 0$, thus the entanglement is an increasing function of the squeezing parameter $r$.

Next, we specifically identify the quadratures which are squeezed. The annihilation operators $\hat{a}_\phi,\hat{b}_\phi$ of the two modes in state $|\phi\rangle$, which is related to the vacuum state by the unitary transformation given in (\ref{squeezed}), are
\begin{numparts}
\begin{eqnarray}
\label{aphi}\hat{a}_\phi &=e^{-r(\hat{a}^+\hat{b}^+-\hat{a}\hat{b})}\hat{a}e^{r(\hat{a}^+\hat{b}^+-\hat{a}\hat{b})}=\hat{a}\cosh{r}+\hat{b}^+\sinh{r},\\
\label{bphi}\hat{b}_\phi &=e^{-r(\hat{a}^+\hat{b}^+-\hat{a}\hat{b})}\hat{b}e^{r(\hat{a}^+\hat{b}^+-\hat{a}\hat{b})}=\hat{a}^+\sinh{r}+\hat{b}\cosh{r},
\end{eqnarray}
\end{numparts}
where the last terms in the above equations are obtained from the Taylor expansion of the middle terms. Now, if we denote the quadrature observables of a mode $\hat{a}$ with
\begin{equation}\label{quadratures}
\hat{X}_a=\frac{\hat{a}+\hat{a}^+}{\sqrt{2}},\quad \hat{P}_a=\frac{\hat{a}-\hat{a}^+}{i\sqrt{2}}
\end{equation}
and define
\begin{equation}\label{quadraturesphi}
\hat{X}^{\phi}=\frac{\hat{X}_{a_\phi}-\hat{X}_{b_\phi}}{\sqrt{2}},\quad \hat{\mathcal{P}}^{\phi}=\frac{\hat{P}_{a_\phi}+\hat{P}_{b_\phi}}{\sqrt{2}},
\end{equation}
where $\hat{X}_{a_\phi},\hat{P}_{a_\phi}$ and $\hat{X}_{b_\phi},\hat{P}_{b_\phi}$ are the quadrature observables corresponding to $\hat{a}_\phi,\hat{b}_\phi$, respectively, then it is not hard to verify that
\begin{equation}
\hat{X}^{\phi}=\frac{e^{-r}}{\sqrt{2}}(\hat{X}_a-\hat{X}_b),\quad \hat{\mathcal{P}}^{\phi}=\frac{e^{-r}}{\sqrt{2}}(\hat{P}_a+\hat{P}_b)
\end{equation}
From the last relation we find
\begin{numparts}
\begin{eqnarray}
\label{variancex}\langle\Delta(\hat{X}^{\phi})^2\rangle &=\frac{e^{-2r}}{2}[\Delta\hat{X}_a^2+\Delta\hat{X}_b^2]=\frac{e^{-2r}}{2},\\
\label{variancep}\langle\Delta(\hat{\mathcal{P}}^{\phi})^2\rangle &=\frac{e^{-2r}}{2}[\Delta\hat{P}_a^2+\Delta\hat{P}_b^2]=\frac{e^{-2r}}{2},
\end{eqnarray}
\end{numparts}
since $\Delta\hat{X}_a^2=\Delta\hat{X}_b^2=\Delta\hat{P}_a^2=\Delta\hat{P}_b^2=1/2$ in the vacuum state. Thus, $\hat{X}^{\phi}$ and $\hat{\mathcal{P}}^{\phi}$ are squeezed, while note that
\begin{equation}\label{commutatorphi}
[\hat{X}^{\phi},\hat{\mathcal{P}}^{\phi}]=0,
\end{equation}
and this is why we used the calligraphic symbol in $\hat{\mathcal{P}}^{\phi}$.

As we explain in the next section, state (\ref{squeezed}) cannot be generated from the initial vacuum state under evolution (\ref{Liouville}), in the blue-detuned regime. For this reason, we consider the following modified target state at the final time $t=T$
\begin{equation}\label{target}
|\psi(T)\rangle=e^{-i\frac{\pi}{4}(\hat{a}^+\hat{a}+\hat{b}^+\hat{b})}|\phi\rangle=\sum_{n=0}^{\infty}\frac{1}{\cosh{r}}\tanh^n{r}e^{-in\frac{\pi}{2}}|n\rangle_a|n\rangle_b.
\end{equation}
The annihilation operators $\hat{a}_\psi,\hat{b}_\psi$ of the two modes in state $|\psi(T)\rangle$ are
\begin{numparts}
\begin{eqnarray}
\label{apsi}\hat{a}_\psi &=e^{i\frac{\pi}{4}(\hat{a}^+\hat{a}+\hat{b}^+\hat{b})}\hat{a}_\phi e^{-i\frac{\pi}{4}(\hat{a}^+\hat{a}+\hat{b}^+\hat{b})}=e^{-i\frac{\pi}{4}}\hat{a}_\phi,\\
\label{bpsi}\hat{b}_\psi &=e^{i\frac{\pi}{4}(\hat{a}^+\hat{a}+\hat{b}^+\hat{b})}\hat{b}_\phi e^{-i\frac{\pi}{4}(\hat{a}^+\hat{a}+\hat{b}^+\hat{b})}=e^{-i\frac{\pi}{4}}\hat{b}_\phi,
\end{eqnarray}
\end{numparts}
where the last terms in the above equations are obtained from the Taylor expansion of the middle terms. If we define
\begin{equation}\label{squeezedquadrature}
\hat{X}^{\psi}=\frac{\hat{X}_{a_\psi}+\hat{P}_{b_\psi}}{\sqrt{2}}, \quad \hat{\mathcal{P}}^{\psi}=\frac{\hat{X}_{b_\psi}+\hat{P}_{a_\psi}}{\sqrt{2}}
\end{equation}
where $\hat{X}_{a_\psi},\hat{X}_{b_\psi},\hat{P}_{a_\psi},\hat{P}_{b_\psi}$ are given by (\ref{quadratures}) for $\hat{a}_\psi,\hat{b}_\psi$, then it is not hard to verify, using (\ref{apsi}), (\ref{bpsi}) and (\ref{quadraturesphi}), that
\begin{equation}
\hat{X}^{\psi}=\frac{\hat{X}^{\phi}+\hat{\mathcal{P}}^{\phi}}{\sqrt{2}}, \quad \hat{\mathcal{P}}^{\psi}=\frac{-\hat{X}^{\phi}+\hat{\mathcal{P}}^{\phi}}{\sqrt{2}}.
\end{equation}
Thus
\begin{equation}
\label{supression}
\langle\Delta(\hat{X}^{\psi})^2\rangle=\langle\Delta(\hat{\mathcal{P}}^{\psi})^2\rangle=\frac{1}{2}[\Delta(\hat{X}^{\phi})^2+\Delta(\hat{\mathcal{P}}^{\phi})^2]=\frac{e^{-2r}}{2}
\end{equation}
and
\begin{equation}\label{commutatorpsi}
[\hat{X}^{\psi},\hat{\mathcal{P}}^{\psi}]=0,
\end{equation}
where we have used (\ref{variancex}), (\ref{variancep}) and (\ref{commutatorphi}). In the target state (\ref{target}) the squeezed quadratures are $\hat{X}^{\psi}$ and $\hat{\mathcal{P}}^{\psi}$, defined in (\ref{squeezedquadrature}).

In order to motivate why we would like to generate such a state, we close this section by briefly reminding its role in the optomechanical teleportation protocol \cite{Hofer11}. According to this protocol, a first light pulse is entangled with the mechanical motion of the mirror forming state $|\psi(T)\rangle$ given in (\ref{target}), while a second light pulse, prepared in the state $|\psi'\rangle$ which will be teleported, interacts with the first pulse in a beam splitter. Two homodyne detectors at the output ports of the beam splitter measure the quadratures $\hat{q}_X=\hat{P}_{a_\psi}+\hat{X}_{a_{\psi'}}$, $\hat{q}_P=\hat{X}_{a_\psi}+\hat{P}_{a_{\psi'}}$. Subsequently, the mirror is displaced in position and momentum by the outcome of these measurements. This feedback brings the mirror to a state $|f\rangle$, with position and momentum given by
\begin{eqnarray*}
\label{mX} \hat{X}_{b_f} &=\hat{X}_{b_\psi}+\hat{q}_X=\hat{X}_{a_{\psi'}}+(\hat{X}_{b_\psi}+\hat{P}_{a_\psi})=\hat{X}_{a_{\psi'}}+\sqrt{2}\hat{\mathcal{P}}^{\psi},\\
\label{mP} \hat{P}_{b_f} &=\hat{P}_{b_\psi}+\hat{q}_P=\hat{P}_{a_{\psi'}}+(\hat{X}_{a_\psi}+\hat{P}_{b_\psi})=\hat{P}_{a_{\psi'}}+\sqrt{2}\hat{X}^{\psi}.
\end{eqnarray*}
For large values of the squuzing parameter $r\rightarrow\infty$ the terms proportional to $\hat{X}^{\psi}, \hat{\mathcal{P}}^{\psi}$ are suppressed and $\hat{X}_{b_f}\rightarrow\hat{X}_{a_{\psi'}}$, $\hat{P}_{b_f}\rightarrow\hat{P}_{a_{\psi'}}$, in other words the state $|\psi'\rangle$ is teleported to the mirror.


\section{Entanglement creation as an optimal control problem in hyperbolic space $H^3$}

\label{sec:problem}

For the evolution described by (\ref{Liouville}), a closed set of equations can be obtained for the second moments of the creation and annihilation operators of the two resonators, for example $\hat{a}^+\hat{a}, \hat{b}^+\hat{b}, \hat{a}^+\hat{b}, \hat{a}^2$ etc., see \cite{Jacobs11}. In order to formulate entanglement generation as a control problem, we do not follow the standard approach to use directly the moment operators \cite{Jacobs11}, but rather use specific linear combinations of them, a set of ten operators which are the generators of the symplectic group $Sp(4)$ \cite{Kim16}, introduced by Dirac \cite{Dirac63}
\begin{eqnarray}\label{operators}
\hat{J}_0&=\frac{1}{2}(\hat{a}^+\hat{a}+\hat{b}\hat{b}^+),\quad\hat{J}_1=\frac{1}{2}(\hat{a}^+\hat{a}-\hat{b}^+\hat{b})\nonumber\\
\hat{J}_2&=\frac{1}{2}(\hat{a}^+\hat{b}+\hat{a}\hat{b}^+),\quad\hat{J}_3=\frac{1}{2i}(\hat{a}^+\hat{b}-\hat{a}\hat{b}^+)\nonumber\\
\hat{K}_1&=\frac{1}{2}(\hat{a}^+\hat{b}^++\hat{a}\hat{b}),\quad\hat{Q}_1=\frac{i}{2}(\hat{a}^+\hat{b}^+-\hat{a}\hat{b})\nonumber\\
\hat{K}_2&=-\frac{1}{4}[(\hat{a}^+)^2+\hat{a}^2-(\hat{b}^+)^2-\hat b^2],\;\hat{K}_3=\frac{i}{4}[(\hat{a}^+)^2-\hat{a}^2+(\hat{b}^+)^2-\hat b^2],\nonumber\\
\hat{Q}_2&=-\frac{i}{4}[(\hat{a}^+)^2-\hat{a}^2-(\hat{b}^+)^2+\hat b^2],\;\hat{Q}_3=-\frac{1}{4}[(\hat{a}^+)^2+\hat{a}^2+(\hat{b}^+)^2+\hat b^2].
\end{eqnarray}
The matrix representation of these operators in terms of Pauli matrices can be found in \cite{Han98}. They satisfy the following commutation relations
\begin{eqnarray}\label{commutators}
  [\hat{J}_i,\hat{J}_j]&=i\epsilon_{ijk}\hat{J}_k,\quad [\hat{J}_i,\hat{K}_j]=i\epsilon_{ijk}\hat{K}_k,\quad[\hat{J}_i,\hat{Q}_j]=i\epsilon_{ijk}\hat{Q}_k\nonumber\\
  \null[\hat{K}_i,\hat{Q}_j]&=i\delta_{ij}\hat{J}_0,\quad[\hat{K}_i,\hat{K}_j]=[\hat{Q}_i,\hat{Q}_j]=-i\epsilon_{ijk}\hat{J}_k,\nonumber\\
  \null[\hat{J}_i,\hat{J}_0]&=0,\quad[\hat{K}_i,\hat{J}_0]=i\hat{Q}_i,\quad [\hat{Q}_i,\hat{J}_0]=-i\hat{K}_i.
\end{eqnarray}
where $\epsilon_{ijk}$ is the Levi-Civita symbol, which is $1$ if $(i,j,k)$ is an even permutation of $(1,2,3)$, $-1$ if it is an odd permutation, and $0$ if any index is repeated, while $\delta_{ij}$ is Kronecker's delta.

In terms of operators (\ref{operators}), the optomechanical Hamiltonian $H$ (\ref{Liouville}) can be written as
\begin{equation}\label{hamiltonianfinal}
  H=2[-\hat{J}_1+g(t)(\hat{K}_1+\hat{J}_2)].
\end{equation}
The expectation value of an operator $\hat{O}$ is $O=\langle\hat{O}\rangle=\mbox{Tr}(\rho\hat{O})$ and, using expression (\ref{hamiltonianfinal}) for the Hamiltonian
in the Liouville-von Neumann equation (\ref{Liouville}) and the commutation relations (\ref{commutators}), we obtain the following systems for the expectation values of the above operators
\begin{equation}\label{4d}
  \left(\begin{array}{c}
    \dot{Q}_1\\
    \dot{Q}_2\\
    \dot{Q}_3\\
    \dot{J}_0
  \end{array}\right)
  =
  \left(\begin{array}{cccc}
    0 & 0 & 2g(t) & -2g(t) \\
    0 & 0 & 2 & 0 \\
    -2g(t) & -2 & 0 & 0 \\
    -2g(t) & 0 & 0 & 0
  \end{array}\right)
  \left(\begin{array}{c}
    Q_1\\
    Q_2\\
    Q_3\\
    J_0
  \end{array}\right),
\end{equation}
\begin{equation}\label{6d}
 \fl\left(\begin{array}{c}
    \dot{K}_1 \\
    \dot{K}_2 \\
    \dot{K}_3 \\
    \dot{J}_1 \\
    \dot{J}_2 \\
    \dot{J}_3
  \end{array}\right)
  =
  \left(\begin{array}{cccccc}
    0 & 0 & 2g(t) & 0 & 0 & 0 \\
    0 & 0 & 2 & 0 & 0 & 2g(t) \\
    -2g(t) & -2 & 0 & 0 & -2g(t) & 0 \\
    0 & 0 & 0 & 0 & 0 & 2g(t) \\
    0 & 0 & -2g(t) & 0 & 0 & 2 \\
    0 & 2g(t) & 0 & -2g(t) & -2 & 0
  \end{array}\right)
  \left(\begin{array}{c}
    K_1 \\
    K_2 \\
    K_3 \\
    J_1 \\
    J_2 \\
    J_3
  \end{array}\right).
\end{equation}

We next move to express the initial and target states of the above systems. Recall that we start from the vacuum state (\ref{initial}),
thus the corresponding initial conditions for the expectation values of the operators defined in (\ref{operators}) are
\numparts\begin{eqnarray}
\label{initial_conditions_a}
\left(\begin{array}{cccc}
    Q_1 & Q_2 & Q_3 & J_0
\end{array}\right)
=
\left(\begin{array}{cccc}
    0 & 0 & 0 & \frac{1}{2}
\end{array}\right)\\
\label{initial_conditions_b}
\left(\begin{array}{cccccc}
    K_1 & K_2 & K_3 & J_1 & J_2 & J_3
\end{array}\right)
=
\left(\begin{array}{cccccc}
    0 & 0 & 0 & 0 & 0 & 0
\end{array}\right).
\end{eqnarray}\endnumparts
The target entangled state is (\ref{target}) and using (\ref{apsi}), (\ref{bpsi}) and (\ref{aphi}), (\ref{bphi}) we find the following expectation values of operators (\ref{operators}) in this state
\begin{eqnarray}\label{final_conditions}
\left(\begin{array}{cccc}
    Q_1 & Q_2 & Q_3 & J_0
\end{array}\right)
=
\left(\begin{array}{cccc}
    -\frac{\sinh{2r}}{2} & 0 & 0 & \frac{\cosh{2r}}{2}
\end{array}\right)\nonumber\\
\left(\begin{array}{cccccc}
    K_1 & K_2 & K_3 & J_1 & J_2 & J_3
\end{array}\right)
=
\left(\begin{array}{cccccc}
    0 & 0 & 0 & 0 & 0 & 0
\end{array}\right).\nonumber
\end{eqnarray}

Now observe from (\ref{initial_conditions_b}) that the initial values of $K_1, K_2, K_3, J_1, J_2, J_3$ are zero. System (\ref{6d}) is linear in these variables and homogeneous, thus these expectation values remain zero during the whole time evolution. Consequently, when starting from the vacuum state, the generation of the target entangled state is solely described by system (\ref{4d}). If we normalize the expectation values $Q_1, Q_2, Q_3, J_0$ with $\frac{1}{2}$, but keep the same notation for these variables, then system (\ref{4d}) remains unchanged, while the starting and target points become
\numparts\begin{eqnarray}\label{start}
\left(\begin{array}{cccc}
    Q_1 & Q_2 & Q_3 & J_0
\end{array}\right)
=
\left(\begin{array}{cccc}
    0 & 0 & 0 & 1
\end{array}\right)\\
\label{end}
\left(\begin{array}{cccc}
    Q_1 & Q_2 & Q_3 & J_0
\end{array}\right)
=
\left(\begin{array}{cccc}
    -\sinh{2r} & 0 & 0 & \cosh{2r}
\end{array}\right).
\end{eqnarray}\endnumparts
Using the initial condition (\ref{start}) and Eq. (\ref{4d}) we find that the system variables satisfy
\begin{equation}\label{hyperbolic}
Q_1^2+Q_2^2+Q_3^2-J_0^2 = -1,
\end{equation}
which is the equation of a hyperboloid of two sheets. But from the definition of $\hat{J}_0$ in (\ref{operators}) it is obviously $J_0>0$, thus
system (\ref{4d}) actually evolves in the upper sheet, the hyperbolic space $H^3$ \cite{Bengsson98}.

This space is closely related to the Minkowski space. We use the usual notation for contravariant and covariant four-vectors
\numparts\begin{eqnarray}\label{contra}
x^{\mu}=\left(\begin{array}{cccc}
     Q_1 & Q_2 & Q_3 & J_0
\end{array}\right)\\ \label{cov}
x_{\mu}=\left(\begin{array}{cccc}
    Q_1 & Q_2 & Q_3 & -J_0
\end{array}\right)
\end{eqnarray}\endnumparts
and for the scalar product between two four-vectors $a,b$
\begin{equation}\label{products}
a\cdot b= a^{\mu}b_{\mu},
\end{equation}
where the Einstein summation convention over repeated indices is assumed. The geodesic distance $d$ between two points in the space, represented by $a$ and $b$, is given by
\begin{equation}\label{spacelike}
d=\cosh^{-1}{(-a\cdot b)},
\end{equation}
where note that for points in $H^3$ it can be shown that it is always
\begin{equation}
\label{product_range}
a\cdot b\leq -1,
\end{equation}
thus the geodesic distance is well defined.
For the initial vacuum state (\ref{start}) and the final squeezed state (\ref{end}) we have
\begin{equation}
x(0)\cdot x(T)=-\cosh{2r},
\end{equation}
and from (\ref{spacelike}) we find the geodesic distance to be
\begin{equation}\label{geodetic_distance}
d=2r,
\end{equation}
twice the squeezing parameter, an aesthetically pleasant expression.

Up to now we have implicitly assumed that the value of the squeezing parameter $r$ at the final state is given. In such a case, a meaningful question to ask is how should we pick the time-varying coupling $g(t)$, which is bounded due to experimental limitations as below
\begin{equation}\label{bound}
-G\leq g(t)\leq G,\quad G>0,
\end{equation}
in order to reach the target point in minimum time $T$. Although in the subsequent sections we do not actually study this minimum-time problem but a closely related one, here we make some simple observations about it. First of all note that, since the dynamics are governed by system (\ref{4d}), which imposes certain restrictions on the allowed paths and speed, the path corresponding to the shortest geodesic distance may not be permitted or may not be the fastest. Next, consider the quantity
\begin{eqnarray}\label{speed}
\dot{x}^2&=\dot{x}^\mu\dot{x}_\mu=\dot{Q}_1^2+\dot{Q}_2^2+\dot{Q}_3^2-\dot{J}_0^2\nonumber\\
         &=4\Big\{g^2(t)(Q_3-J_0)^2+2g(t)Q_1Q_2+Q_2^2+Q_3^2\Big\}.
\end{eqnarray}
With some algebra we can show that $\dot{x}^2\geq 0$ for points in $H^3$, and note that there is no contradiction with (\ref{product_range}), since $\dot{x}$ belongs to the tangent space of $H^3$. Since the quantity $\sqrt{\dot{x}^2}$ actually corresponds to the speed in the flat $(3+1)$ Minkowski space with metric tensor $\mbox{diag}(1, 1, 1,-1)$, where $H^3$ is embedded, it is in principle desirable to move along paths where this quantity is maximized. But from (\ref{speed}) observe that $\dot{x}^2$ is actually a quadratic function of the control $g(t)$, with the coefficient of $g^2(t)$ being positive
\begin{equation*}
(Q_3-J_0)^2>0,
\end{equation*}
since $Q_3=J_0$ would imply in (\ref{hyperbolic}) $Q_1^2+Q_2^2=-1$, which is obviously not true. Thus the instantaneous speed $\dot{x}^2$ is a convex function of the coupling, which is restricted as in (\ref{bound}), and is maximized when $g(t)$ takes values at the boundaries $\pm G$. Note of course that, since the choice of $g(t)$ also affects the path travelled, there might be cases where the minimum-time path is not the one along which the speed is maximized.

Inspired from the above analysis and the well known fact that optomechanical state swap can be achieved for specific constant values of the coupling rate and the duration $T$ \cite{Stefanatos16}, it is tempting to examine whether a constant coupling $g(t)=g$ can transfer the system to the desired target state. We show that, unfortunately, this is not the case for squeezing. It can be verified using the system equation (\ref{4d}) that for \emph{constant} $g(t)$ the quantity $\dot{x}^2$ given in (\ref{speed}) is also constant. But using the initial and final conditions (\ref{start}) and (\ref{end}) we obtain
\begin{equation*}
\dot{x}^2(T)=4g^2\cosh^2{2r}\geq 4g^2=\dot{x}^2(0),
\end{equation*}
where the equality holds only for $r=0$. The conclusion is that in order to reach the target state the coupling has to change in time. 

The problem that we actually study in the rest of the paper is closely related to the minimum-time problem stated above. Specifically, we fix the duration $T$ and ask what is the maximum value of the squeezing parameter, corresponding to the maximum entanglement, which can be obtained at the final time $t=T$. In order to precisely formulate this problem and also reduce the dimension of the system from four to three, we use the constant of the motion (\ref{hyperbolic}). Recall that $J_0>0$, thus
\begin{equation}
\label{J0}
J_0=\sqrt{1+Q_1^2+Q_2^2+Q_3^2}.
\end{equation}
Using this relation we end up with the three-dimensional system
\begin{numparts}
\begin{eqnarray}
\label{Q1}\dot{Q}_1&=-2g\left(\sqrt{1+Q_1^2+Q_2^2+Q_3^2}-Q_3\right)\\
\label{Q2}\dot{Q}_2&=2Q_3\\
\label{Q3}\dot{Q}_3&=-2gQ_1-2Q_2.
\end{eqnarray}
\end{numparts}
Starting from
\begin{equation}
\label{initial_point}
\left(\begin{array}{ccc}
    Q_1 & Q_2 & Q_3
\end{array}\right)
=
\left(\begin{array}{ccc}
    0 & 0 & 0
\end{array}\right),
\end{equation}
we would like to find the time-varying control $g(t)$, bounded as in (\ref{bound}), which minimizes $Q_1(T)$ while it drives the other two state variables to the final conditions
\begin{equation}
\label{finalQ}
Q_2(T)=Q_3(T)=0.
\end{equation}
Note that, since $Q_1(T)=-\sinh{2r}$, its minimization corresponds to maximizing the squeezing parameter $r$ and the entanglement.

We finally show that for two durations $T,T'$, with $T'>T$, the corresponding minimum values of $Q_1$ satisfy $Q_1(T')\leq Q_1(T)$, i.e. the minimum value of $Q_1$ is a nonincreasing function of the final time, thus the squeezing parameter and the entanglement are nondecreasing functions of it. Indeed, if $g(t)$ is the optimal control for $0\leq t\leq T$, which drives the system at the point $(Q_1(T),0,0)$, then the control
\begin{equation}
\label{bang-bang}
g'(t)=\left\{\begin{array}{cl} g(t), & 0\leq t\leq T \\0, & T<t\leq T'\end{array}\right.,
\end{equation}
drives the system at the same point at the final time $t=T'$. Thus, for the larger duration $T'$ we can obtain at least the value $Q_1(T)$, corresponding to the minimum for the smaller duration $T$. Consequently, the minimum corresponding to $T'$ satisfies $Q_1(T')\leq Q_1(T)$. This monotonic behavior of the objective function indicates the dual character of our maximum-entanglement problem and the minimum-time problem stated above. If we solve a series of maximum-entanglement problems for increasing duration $T$, then the solution of the minimum-time problem is the shortest time for which the desired level of entanglement is obtained.

\section{Analysis of the optimal solution}

\label{sec:solution}

In this section we use optimal control theory \cite{Bryson75} to obtain some characteristics of the optimal solution for the maximum-entanglement problem.
The control Hamiltonian for this problem is defined as
\begin{eqnarray}
\label{control_hamiltonian}
H_c&=\lambda_1\dot{Q}_1+\lambda_2\dot{Q}_2+\lambda_3\dot{Q}_3\nonumber\\
   &=2g(-\lambda_1\sqrt{1+Q_1^2+Q_2^2+Q_3^2}+\lambda_1Q_3-\lambda_3Q_1)+2(\lambda_2Q_3-\lambda_3Q_2),
\end{eqnarray}
where the costates $\lambda_i, i=1,2,3$ satisfy the equations
\begin{numparts}
\begin{eqnarray}
\label{lambda1}\dot{\lambda}_1&=-\frac{\partial H_c}{\partial Q_1}=2g\left(\frac{\lambda_1Q_1}{\sqrt{1+Q_1^2+Q_2^2+Q_3^2}}+\lambda_3\right),\\
\label{lambda2}\dot{\lambda}_2&=-\frac{\partial H_c}{\partial Q_2}=2\left(g\frac{\lambda_1Q_2}{\sqrt{1+Q_1^2+Q_2^2+Q_3^2}}+\lambda_3\right),\\
\label{lambda3}\dot{\lambda}_3&=-\frac{\partial H_c}{\partial Q_3}=2\left[g\left(\frac{\lambda_1Q_3}{\sqrt{1+Q_1^2+Q_2^2+Q_3^2}}-\lambda_1\right)-\lambda_2\right].
\end{eqnarray}
\end{numparts}
Since we want to minimize $Q_1(T)$, the following terminal condition should be satisfied by the corresponding costate \cite{Bryson75}
\begin{equation}
\label{lambda1_terminal}
\lambda_1(T)=1.
\end{equation}

According to Pontryagin's Minimum Principle, the optimal control $g(t)$ is chosen such that the control Hamiltonian is minimized for $0\leq t\leq T$. But the control appears only linearly in $H_c$, while it is bounded as in (\ref{bound}), thus the optimal choice of $g$ depends on the sign of the quantity which multiplies it in (\ref{control_hamiltonian}), the so-called \emph{switching function}
\begin{equation}
\label{switching_function}
\Phi=-\lambda_1\left(\sqrt{1+Q_1^2+Q_2^2+Q_3^2}-Q_3\right)-\lambda_3Q_1.
\end{equation}
Specifically, the optimal control is
\begin{equation}
\label{optimal_control}
g(t)=\left\{\begin{array}{cl} G, & \Phi<0 \\\mbox{singular}, & \Phi=0\\-G, & \Phi>0\end{array}\right..
\end{equation}
Observe that when the switching function is nonzero, the optimal control takes the corresponding boundary value which minimizes $H_c$ and we call this a \emph{bang} pulse. When the switching function is zero during a time interval of nonzero measure, the control is called \emph{singular} and is calculated using the derivatives of $\Phi$, as we demonstrate in the next paragraph.

During a time interval where the control is singular, the switching function and its time derivatives are zero
\begin{equation*}
\Phi=\dot{\Phi}=\ddot{\Phi}=\ldots=0.
\end{equation*}
For the problem at hand we find, using the system and costate differential equations
\begin{numparts}
\begin{eqnarray}\
\label{fi}\fl\Phi&=0\Rightarrow\left(\sqrt{1+Q_1^2+Q_2^2+Q_3^2}-Q_3\right)\lambda_1+Q_1\lambda_3=0\\
\label{fidot}\fl\dot{\Phi}&=0\Rightarrow-Q_2\lambda_1+Q_1\lambda_2=0\\
\label{fiddot}\fl\ddot{\Phi}&=0\Rightarrow-Q_3\lambda_1+g\left(Q_3-\sqrt{1+Q_1^2+Q_2^2+Q_3^2}\right)\lambda_2+(Q_1-gQ_2)\lambda_3&=0
\end{eqnarray}
\end{numparts}
Now observe that the costates $\lambda_i$ cannot be simultaneously zero at any moment, since in such case the linear homogeneous system of first order differential equations (\ref{lambda1})-(\ref{lambda3}) for the costates would imply $\lambda_i(t)=0$ for $0\leq t\leq T$, violating the terminal condition (\ref{lambda1_terminal}). This non-triviality of the costates requires the determinant corresponding to the linear homogeneous system (\ref{fi})-(\ref{fiddot}) to be zero, which leads to the condition
\begin{equation}
\label{surface}
Q_1=0.
\end{equation}
When the control is singular, the system evolves in the above surface, which is called \emph{singular surface}. On this surface it is also $\dot{Q}_1=0$, and from system equation (\ref{Q1}) we find $g\left(\sqrt{1+Q_1^2+Q_2^2+Q_3^2}-Q_3\right)\Rightarrow g=0$, since obviously $\sqrt{1+Q_1^2+Q_2^2+Q_3^2}-Q_3>0$. Thus, the singular control is
\begin{equation}
\label{singular_control}
g(t)=0.
\end{equation}
From the system equations it is obvious that a singular trajectory corresponds to a rotation around $Q_1$-axis on the plane $Q_1=0$.

Note that Pontryagin's Minimum Principle provides necessary conditions for optimality. In the case of singular control, there is an additional necessary condition, the so-called generalized Legendre-Clebsch condition \cite{Bryson75}. For the minimization problem at hand it takes the form
\begin{eqnarray}
&(-1)^1\frac{\partial}{\partial g}\left[\frac{d^2}{dt^2}\left(\frac{\partial H_c}{\partial g}\right)\right]\geq 0\nonumber\\
\label{Kelley}\Rightarrow &\lambda_2\left(Q_3-\sqrt{1+Q_1^2+Q_2^2+Q_3^2}\right)-\lambda_3Q_2\leq 0.
\end{eqnarray}
We will use this condition in the next section as an extra test of optimality.

We close this section by observing that, although we have identified the possible values of the optimal control, finding the exact pulse sequence is a difficult problem which depends on the duration $T$ and the control bound $G$. For this reason, in the next section we recourse to numerical optimization in order to obtain the optimal pulse sequences for specific values of these parameters. But before doing so, here we exclude certain pulse sequences and this will lead us to the most frequently encountered optimal control profiles. First of all, recall
that in the previous section we showed that a simple constant pulse cannot drive the system to a final state of the desired form. Next, observe that a bang-singular pulse sequence is forbidden because the initial bang pulse cannot reach the singular surface (\ref{surface}), since during this pulse $\dot{Q}_1$ has constant sign, from (\ref{Q1}) with $g(t)=\pm G$, while $Q_1(0)=0$. The sequence singular-bang is equivalent to a simple bang pulse, since the singular control is $g(t)=0$ and the system remains at the starting point (the origin) during this part, thus it is also excluded. We examine next a bang-bang sequence, composed by two bang pulses ($\pm G\mp G$). Observe that, when the duration $T$ is fixed, there is only one free parameter, the duration of one of the pulses, while there are two final conditions (\ref{finalQ}) to be satisfied. Thus, such a pulse sequence cannot in general be optimal and this may happen only for certain values of $T$ and $G$. In order to satisfy the two final conditions (\ref{finalQ}), a sequence of at least three pulses is necessary, where the durations of the first two can be taken as free parameters for fixed $T$. Indeed, in the next section we present a numerical example where the optimal control is a bang-bang-bang sequence. Sequences of the form bang-bang-singular, singular-bang-bang and bang-singular-bang can be excluded using similar reasoning as above. The simplest pulse sequence containing a singular arc has the form bang-bang-singular-bang and this is the optimal pulse sequence that we encountered in most of our simulations.

\section{Numerical examples and discussion}

\label{sec:examples}

In this section, we use the freely available optimal control solver BOCOP \cite{bocop} to obtain numerically the optimal solutions for several illustrative examples. In the BOCOP software package, the continuous-time optimal control problem is approximated by a finite-dimensional optimization problem, using time discretization. The resultant nonlinear programming problem is subsequently solved using the nonlinear solver Ipopt.

\begin{figure*}[t!]
 \centering
		\begin{tabular}{cc}
     	\subfigure[$\ $Optimal control, $G=1, T=\pi$]{
	            \label{fig:control1}
	            \includegraphics[width=.45\linewidth]{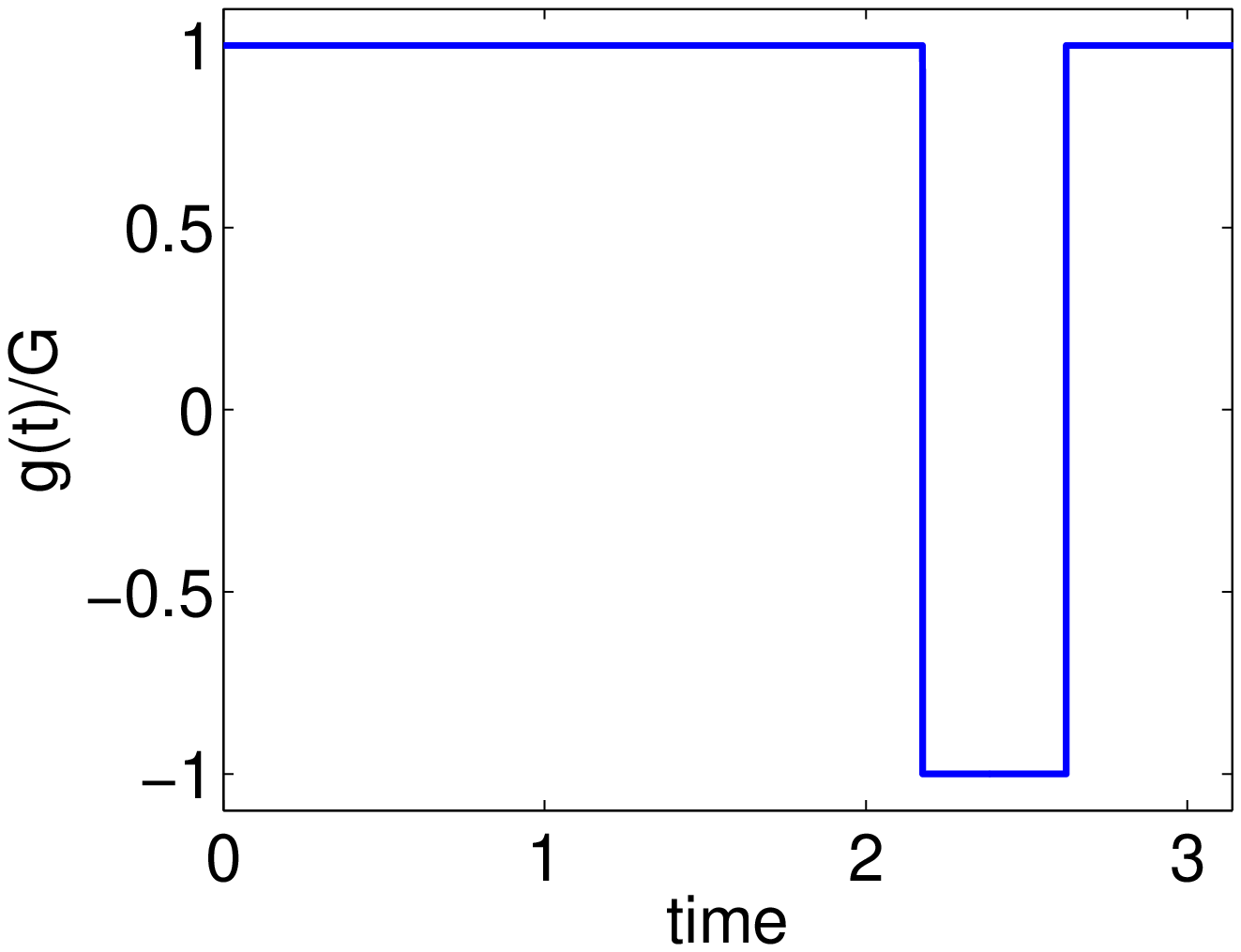}} &
        \subfigure[$\ $Optimal control, $G=2, T=\pi/2$]{
	            \label{fig:control2}
	            \includegraphics[width=.45\linewidth]{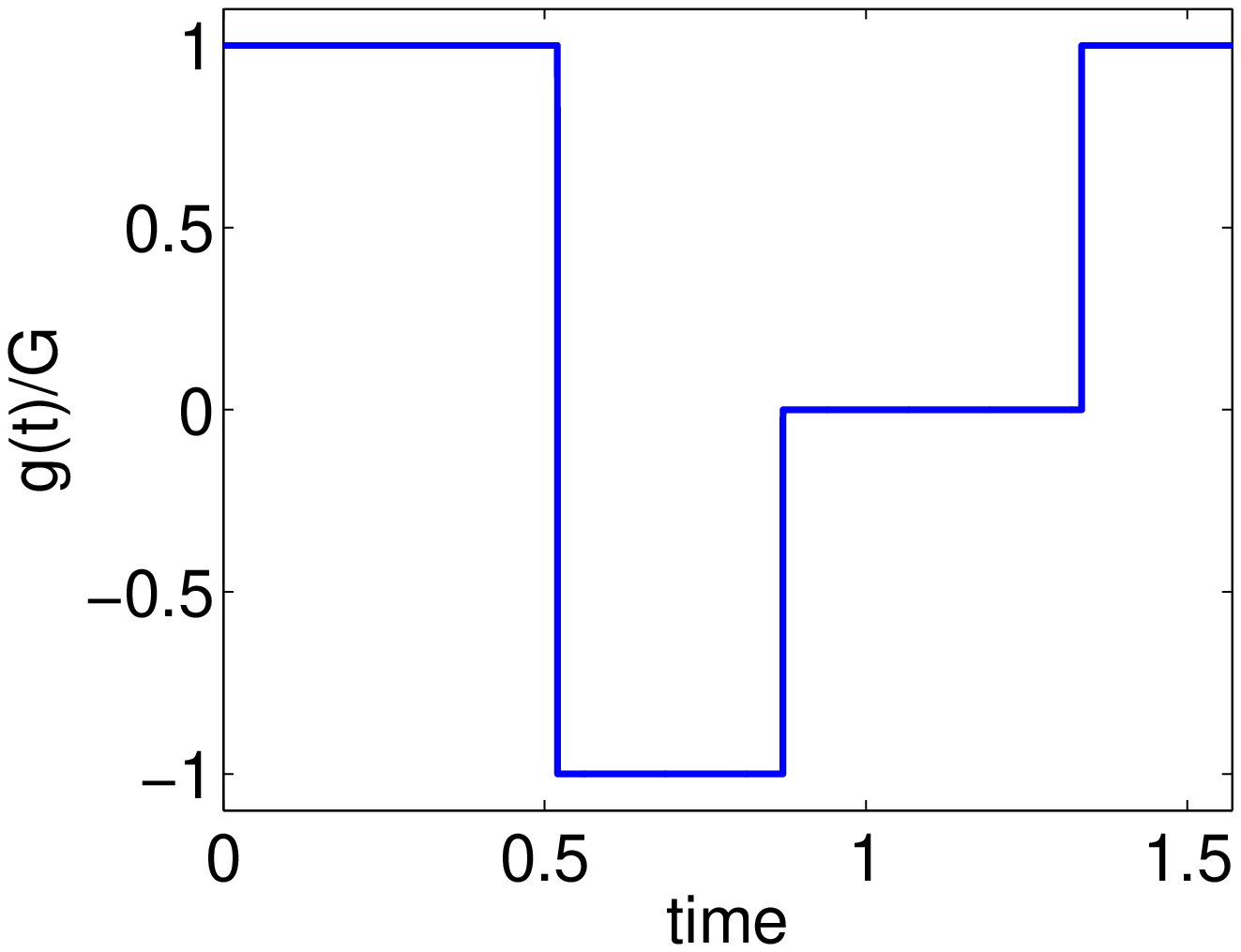}} \\
	    \subfigure[$\ $Swithcing function, $G=1, T=\pi$]{
	            \label{fig:switching1}
	            \includegraphics[width=.45\linewidth]{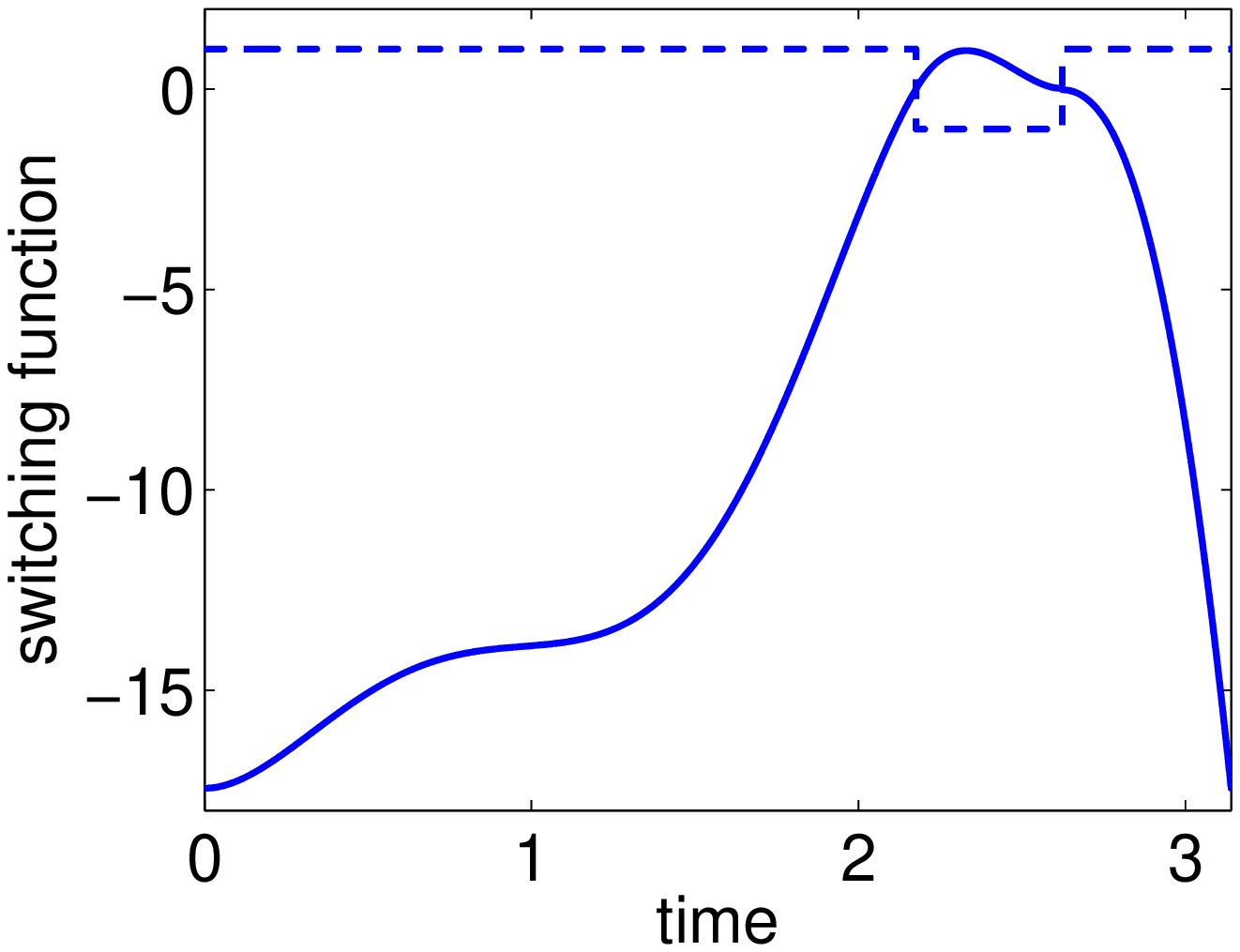}} &
        \subfigure[$\ $Optimality tests, $G=2, T=\pi/2$]{
	            \label{fig:switching2}
	            \includegraphics[width=.45\linewidth]{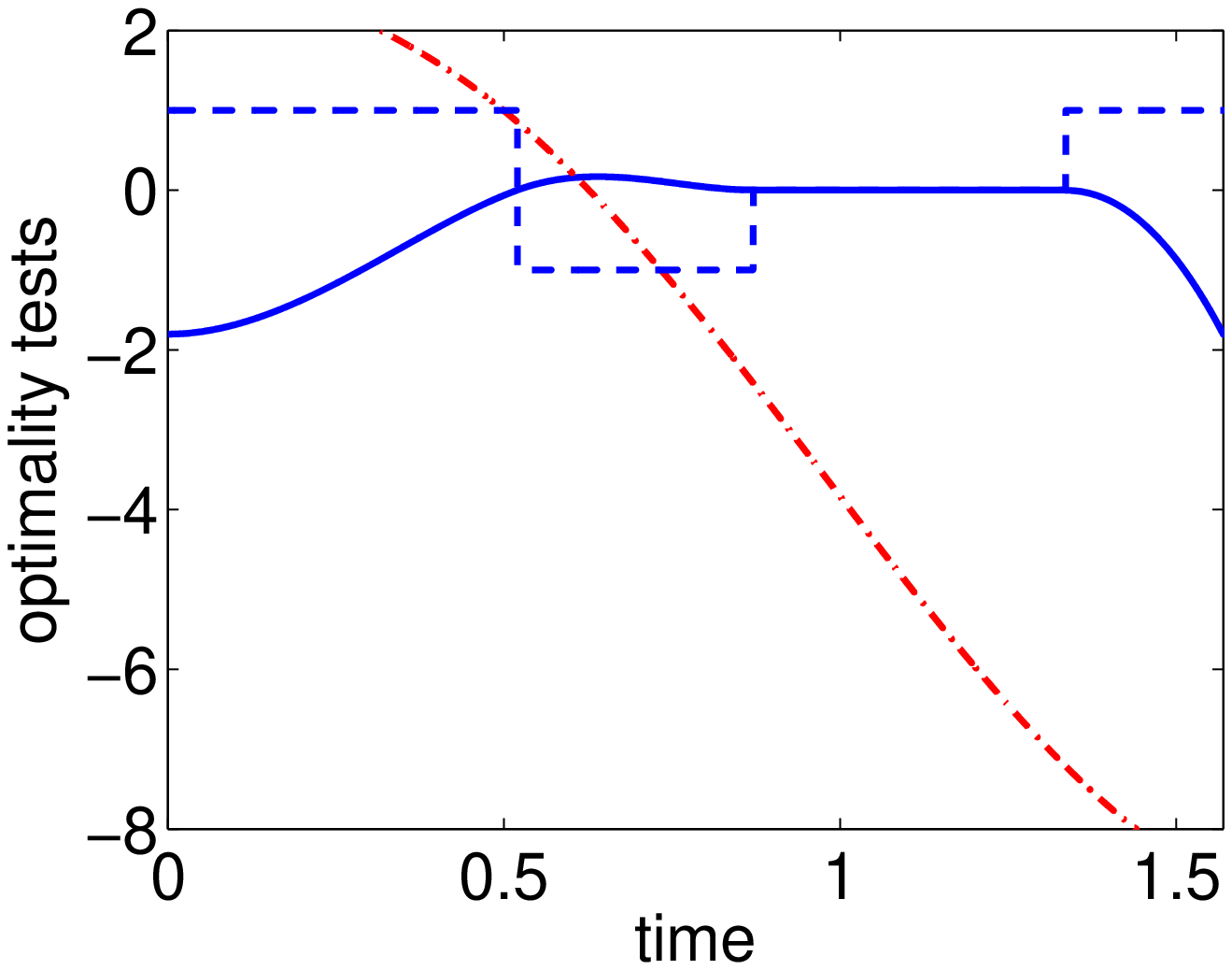}} \\
        \subfigure[$\ $Optimal trajectory, $G=1, T=\pi$]{
	            \label{fig:trajectory1}
	            \includegraphics[width=.45\linewidth]{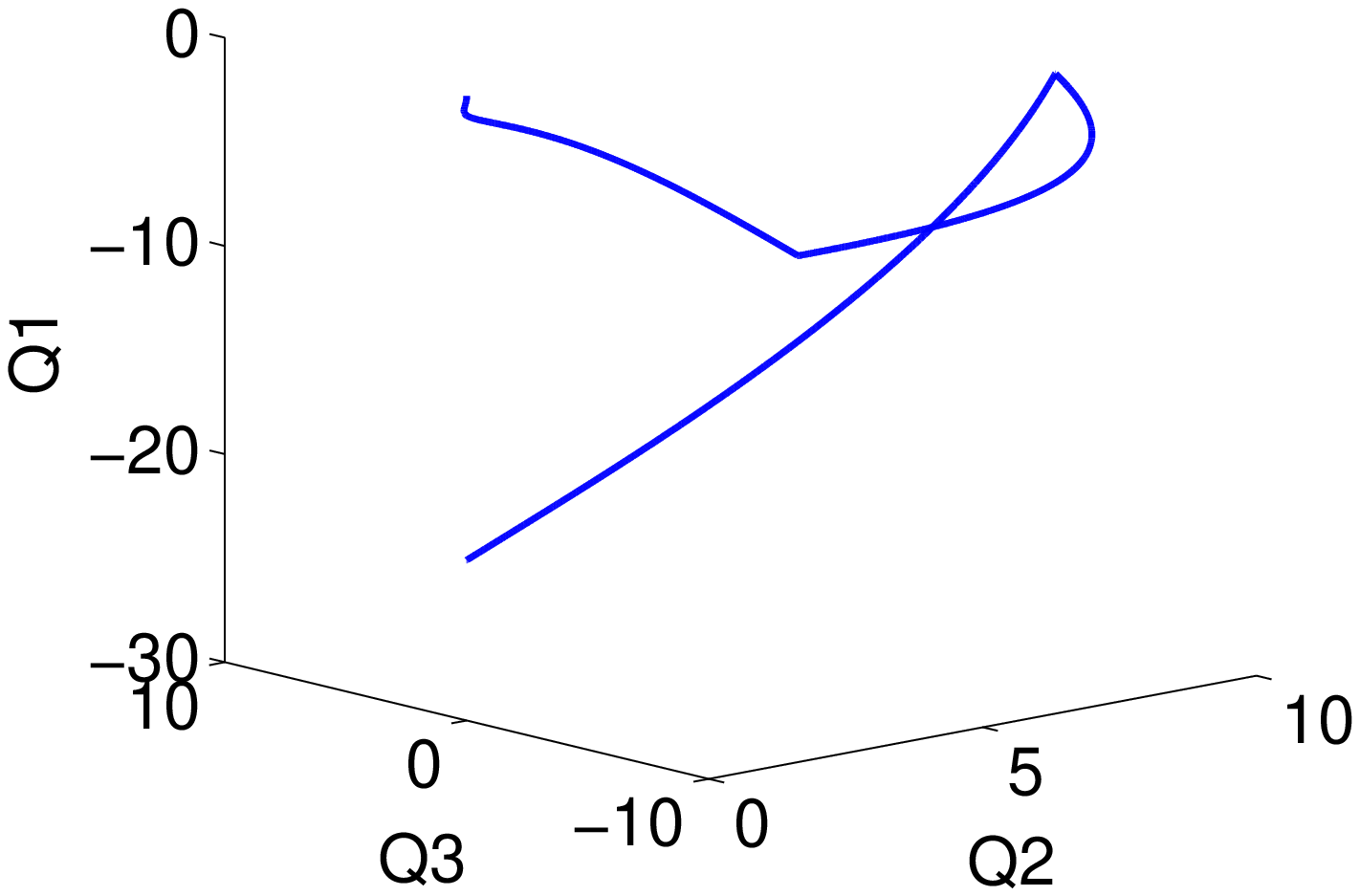}} &
        \subfigure[$\ $Optimal trajectory, $G=2, T=\pi/2$]{
	            \label{fig:trajectory2}
	            \includegraphics[width=.45\linewidth]{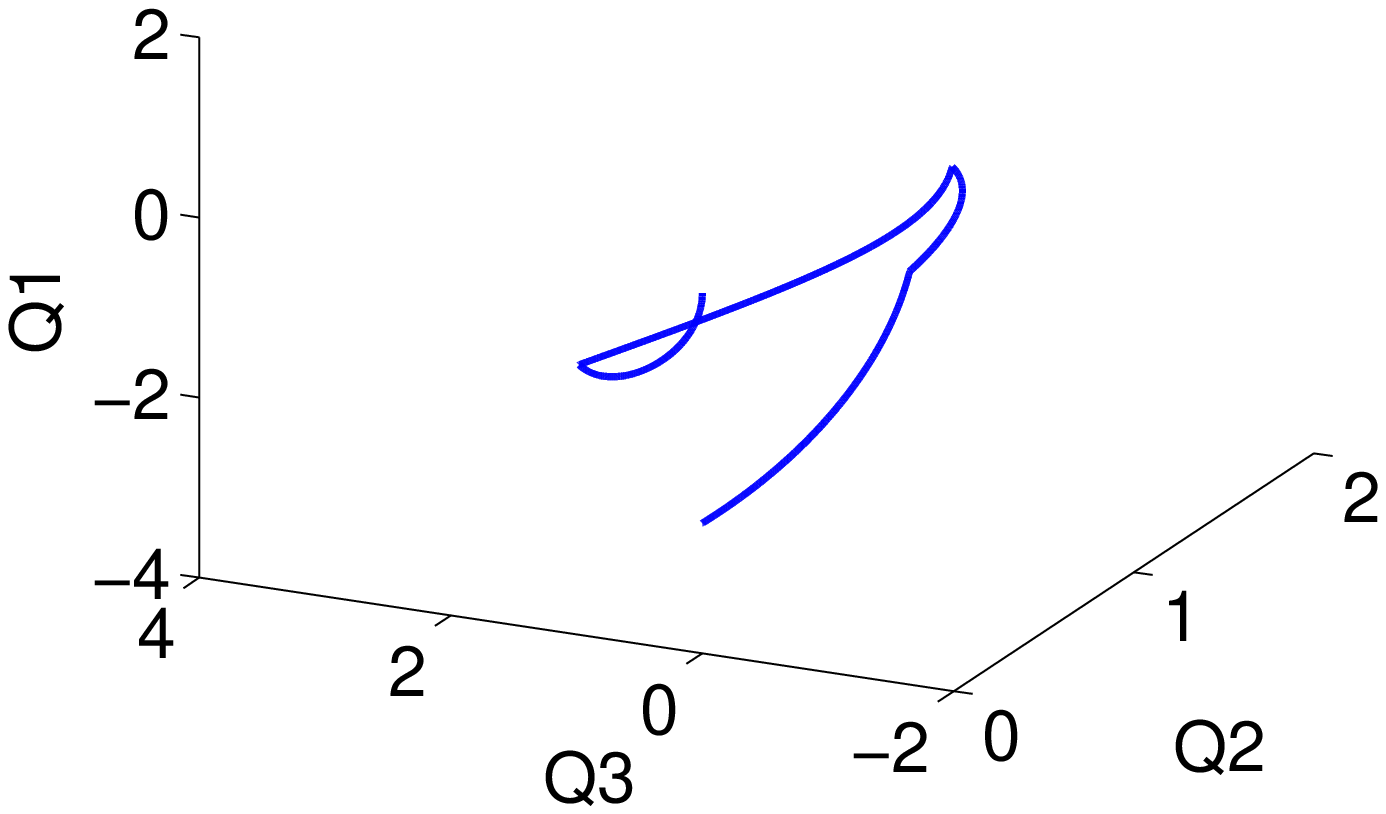}}
		\end{tabular}
\caption{(Color online) Normalized optimal control $g(t)/G$ for (a) $G=1, T=\pi$ and (b) $G=2, T=\pi/2$. (c) Switching function (blue solid line) superimposed on the optimal control (blue dashed line), for $G=1, T=\pi$. (d) Switching function (blue solid line) and generalized Legendre-Clebsch condition (red dashed-dotted line) superimposed on the optimal control (blue dashed line), for $G=2, T=\pi/2$. Optimal trajectories for (e) $G=1, T=\pi$ and (f) $G=2, T=\pi/2$.}
 \label{fig:examples}
\end{figure*}

In the first example that we consider we take $G=1$, a maximum coupling equal to the frequency of the mechanical resonator, and $T=\pi$, which is half the period of this oscillator. We use a time discretization of 20000 points. In Fig. \ref{fig:control1} we plot the numerically obtained normalized optimal control $g(t)/G$. Observe that it has the form bang-bang-bang. The corresponding maximum value of the squeezing parameter is $r=1.8990$. The accuracy for the boundary constraints is of the order of $10^{-28}$, while that for the dynamic constraints (the discretized system equations) of the order of $10^{-15}$. In Fig. \ref{fig:switching1} we plot the corresponding switching function (\ref{switching_function}) (blue solid line) superimposed on the optimal control (blue dashed line). Observe that $g=G$ when $\Phi <0$ and $g=-G$ when $\Phi >0$, in accordance to the Minimum Principle (\ref{optimal_control}). In Fig. \ref{fig:trajectory1} we plot the optimal trajectory which consists of three segments, each corresponding to a bang control.

The next case that we consider is with $G=2$ and $T=\pi/2$. Again, we use a time discretization of 20000 points. In Fig. \ref{fig:control2} we plot the normalized optimal control, which now has the form bang-bang-singular-bang. The maximum value of the squeezing parameter is $r=0.8336$, with accuracies for boundary and dynamic constraints of the order of $10^{-29}$ and $10^{-16}$, respectively. Observe that this accuracy is better than in the previous example, since we use the same discretization for a shorter time interval. In Fig. \ref{fig:switching2} we plot the switching function (blue solid line) and the left hand side of the generalized Legendre-Clebsch condition (\ref{Kelley}) (red dadhed-dotted line), superimposed on the optimal control (blue dashed line). Observe that $g=G$ for $\Phi <0$ and $g=-G$ for $\Phi >0$, as in the previous case, while $g=0$ when $\Phi=0$ (singular arc). Note that on the singular arc the generalized Legendre-Clebsch condition is also satisfied. In Fig. \ref{fig:trajectory2} we plot the optimal trajectory where four segments are clearly distinguished, each corresponding to a constant control interval. Observe that the singular third segment is actually a rotation around $Q_1$-axis on the plane $Q_1=0$. Here we point out that singular optimal controls appear in the context of quantum control in the contrast imaging problem of Nuclear Magnetic Resonance \cite{Bonnard12}, while they are quite common in aerospace applications, see for example the famous Goddard problem of maximizing the altitude of a vertically ascending rocket with a finite amount of fuel \cite{Tsiotras92}.

\begin{figure}[t]
\centering
\includegraphics[width=0.7\linewidth]{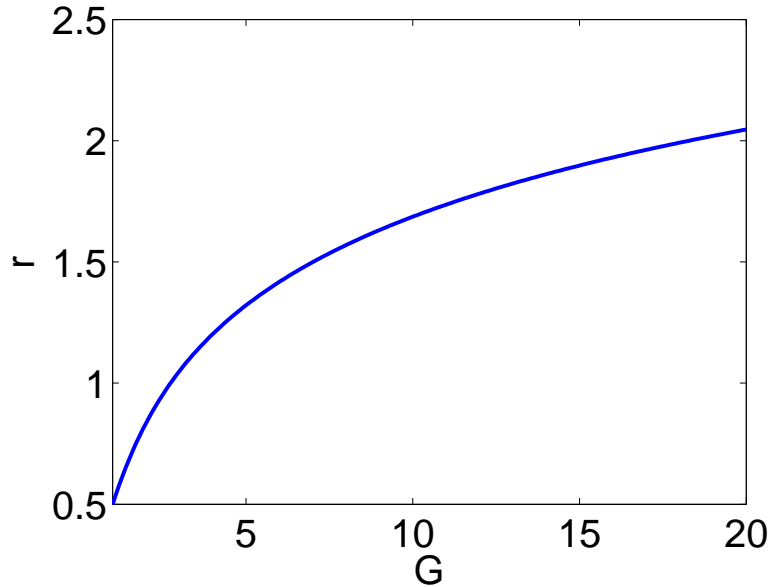}
\caption{Squeezing parameter $r$ as a function of the maximum control amplitude $G$ for fixed $T=\pi/2$.}%
\label{fig:batchG}%
\end{figure}

\begin{figure}[t]
\centering
\includegraphics[width=0.7\linewidth]{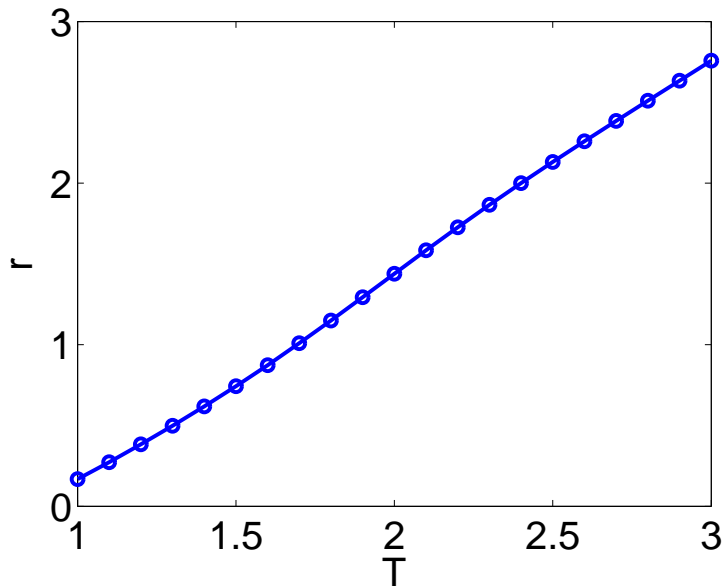}
\caption{Squeezing parameter $r$ as a function of duration $T$ for fixed $G=2$.}%
\label{fig:batchT}%
\end{figure}

In Fig. \ref{fig:batchG} we plot the squeezing parameter $r$ as a function of the control bound $G$, for $1\leq G\leq 20$ with a step $\Delta G=0.1$ and for fixed duration $T=\pi/2$. For all these optimizations we have found that the optimal control has the form bang-bang-singular-bang. The percentage of the time occupied by the singular arc is increasing with increasing $G$. Finally, in Fig. \ref{fig:batchT} we plot the squeezing parameter $r$ as a function of $T$, for $1\leq T\leq 3$ with a step $\Delta T=0.1$ and for fixed control bound $G=2$. Observe that this last plot can be used to find the necessary time to obtain a desired level of squeezing-entanglement. From the last two figures it is obvious that in the strong coupling regime, where $G$ is comparable or larger than the frequency of the mechanical resonator, a substantial amount of entanglement can be created within a fraction of a single oscillation period, which is $2\pi$.

\section{Conclusion}

\label{sec:conclusion}

In this paper, we formulated the creation of optomechanical entanglement as an optimal control problem and used optimal control theory to study the optimal solution. We subsequently used numerical optimization to obtain the optimal pulse sequences for several examples. Although here we considered as the initial state of both oscillators the vacuum, it is our intention to extend the present work to the more practical situation where the mechanical resonator is initially in a thermal state. Note that in this case the problem is complicated by the fact that both systems (\ref{4d}) and (\ref{6d}) have nonzero initial conditions.


\end{document}